\documentstyle{elsart}
\input psfig
\input epsf
\begin{document}

\hfill INP 1829/PH \vspace{5mm}

\begin{frontmatter}
\title{In-Medium Modifications of Hadron Masses and Chemical Freeze-Out
in Ultra-Relativistic Heavy-Ion Collisions $^*$ } 
\thanks{Research supported in part by the Polish State Committee for
        Scientific Research, grant 2P03B-080-12}
\author{Wojciech Florkowski and Wojciech Broniowski}
\address{H. Niewodnicza\'nski Institute of Nuclear Physics, \\
        ul. Radzikowskiego 152,  PL-31342 Krak\'ow, Poland}

\begin{abstract}
We confront the hypothesis of chemical freeze-out in ultra-relativistic
heavy-ion collisions with the hypothesis of large modifications of
hadron masses in nuclear medium. We find that the thermal-model predictions
for the ratios of particle multiplicities are sensitive to the
values of in-medium hadronic masses. In particular, the  $\pi^+/p$ ratio
decreases by 35\% when the masses of all hadrons (except for pseudo-Goldstone
bosons) are scaled down by 30\%.
\end{abstract}
\end{frontmatter}
\vspace{-7mm} PACS: 25.75.Dw, 21.65.+f, 14.40.-n

Recent theoretical studies \cite{BM12,BM3} indicate that hadron yields
and ratios in ultra-relativistic heavy-ion collisions agree with
predictions of a simple thermal model. According to this model all
measured abundances of hadrons are consistent with the assumption of
complete thermalization of hadronic matter at a temperature
$T_{chem}$, baryon chemical potential $\mu _{chem}^B$, strangeness
chemical potential $\mu _{chem}^S$, and isospin chemical potential
$\mu _{chem}^I$.  Thermodynamic parameters obtained from this type of
analysis characterize the point in the evolution of the system when
``primordial'' hadron content is established.  One refers to this
point as to the {\it chemical freeze-out}.  At this stage the system
is a mixture of stable particles (pions, kaons, nucleons,...) and
resonances ($\rho, \omega, \Delta,...$). In the subsequent evolution
the resonances decay, contributing to the final (observed) multiplicities of
stable particles.

It has been shown that the
chemical freeze-out parameters at CERN/SPS, BNL/AGS and GSI/SIS
energies all correspond to a unique value of the energy per hadron
\cite{CR}. Moreover, statistical models
\cite{Bec} are able to
reproduce the particle yields in $e^+ e^-$ collisions. A striking
observation is that at very high energies the temperature
$T_{chem}\sim 170$MeV turns out
to be the same for both elementary and nuclear collisions, although
the final-state hadronic interactions are completely absent in the former
case. This may indicate that chemical equilibrium is {\it pre-established}
by the hadronization process \cite{Bec,H,Bialas}.

In recent studies one distinguishes the chemical freeze-out from the {\it %
thermal} or {\it kinetic freeze-out} \cite{H}. The latter is defined as
the decoupling of strongly interacting matter produced in high-energy
nuclear collisions into a system of essentially free-streaming particles.
After the thermal freeze-out the hadrons practically stop to interact and
travel freely to detectors. Chemical freeze-out is expected to occur at the
same time or before the thermal freeze-out \cite{H}. For Pb+Pb collisions
at CERN/SPS energies one finds that the chemical freeze-out point occurs
significantly earlier than the thermal point. This is indicated by the
measurements of the hadron momentum spectra as well as two-particle momentum
correlations \cite{WH}, which show that the thermal freeze-out
temperature $T_{therm}\sim 100$ MeV is substantially lower than $T_{chem\
}\sim 170$ MeV. In addition, $T_{chem}$ is very close to the expected
critical value for the deconfinement/hadronization phase transition,
therefore the chemical composition of a hadronic system, according to the
presented scenario, should be established shortly after hadronization. 

Since the chemical freeze-out occurs at an early stage of the
evolution of the system, where temperatures and densities are very
high, we expect that hadronic properties, such as masses, widths, or
coupling constants {\em are influenced strongly} by the medium. Such
modifications are predicted in many theoretical calculations
\cite{BR,hatlee,hatsuda,klingl}. Moreover, they seem  much desired
\cite{cassing,li}  in view of the low-mass dilepton enhancement observed in
the CERES \cite{CERES} and HELIOS \cite{HELIOS} experiments. 

In order to study how the chemical freeze-out parameters are affected by the
in-medium change of the hadron masses, we calculate the particle densities 
from the standard ideal-gas equilibrium expression 
\begin{equation}
n_i=\frac{g_i}{2\pi ^2}\int_0^\infty \frac{p^2\ dp}{\exp \left[ \left(
E_i^{*}-\mu _{chem}^BB_i-\mu _{chem}^SS_i-\mu _{chem}^II_i\right)
/T_{chem}\right] \pm 1},  \label{rhoi}
\end{equation}
where $g_i$ is the spin degeneracy factor of the $i$th hadron, $B_i,S_i,I_i$
are its baryon number, strangeness, and the third component of isospin, and $%
E_i^{*}=\sqrt{p^2+\left( m_i^{*}\right) ^2}$  is its energy. The latter
explicitly depends on the in-medium hadron mass $m_i^{*}$. 

In the thermal-model fits \cite{BM12,BM3} one uses Eq. (\ref{rhoi})
with vacuum values of hadron masses $m^*_i=m_i$, and, in addition,
applies finite--size and excluded--volume corrections. They account
for the finite size of the nuclear system and the finite volume
occupied by the individual hadrons. The main effect of such
corrections is a reduction of the absolute yields of particles with
the particle ratios remaining close to predictions of the ideal-gas
approach \cite{BM3}. For Pb+Pb collisions at
CERN/SPS energies the predictions of the thermal model are \cite{BM3}:
\begin{eqnarray}
&&T_{chem}=168\,{\rm MeV},\, \mu _{chem}^B=266{\rm MeV},\,  \nonumber \\
&&\\
&&\mu_{chem}^S=71{\rm MeV},\, \mu _{chem}^I=-5{\rm MeV.} \nonumber
\end{eqnarray}

In our study we use these values in Eq. (\ref{rhoi}) and calculate the
particle densities $n_i$. We take into account all meson resonances
with masses smaller than 1.28 GeV and baryon resonances with masses
smaller than 1.45 GeV. Decays of resonances contribute to the final
(measured) yields of stable hadrons. This is an important effect
\cite{BM12}. The needed branching ratios for the decays of resonances
are taken from experiment \cite{Tablice}. In this paper, for clarity,
we neglect the finite-size and the excluded-volume corrections. As
mentioned above, they do not change significantly the ratios of
particles. Indeed, within our simplified approach we reproduce the
numbers of Ref. \cite{BM3} at the level of 15\%.

In principle, the  in-medium masses of all hadrons may behave differently.
For practical reasons we perform our calculation with the meson and 
baryon masses rescaled by the two universal parameters, $x_M$ and $x_B$,
namely
\begin{equation}
m^*_M = x_M \, m_M, \hspace{0.5cm}  m^*_B = x_B \, m_B.
\end{equation}
In this way we change the masses of all hadrons except for
pseudo-Goldstone bosons, {\it i.e.}, pions, kaons and the eta, whose
masses are kept fixed at vacuum values. In fact, results of many model
calculations show stability of the pion  mass against the
change of temperature and density up to the point where the chiral
phase transition occurs \cite{RKH}. The branching ratios are kept at the
vacuum values.

In Fig. 1 we show the $\pi ^{+}/p$ ratio calculated for different
values of the parameters $x_M$ and $x_B$. The solid line represents
the case when the meson and baryon masses are rescaled the same way,
$x_M=x_B=x$ (BR-scaling \cite{BR}), the dashed line corresponds to the
case when only the baryon masses are changed, $x_M=1$ and $x_B=x$, and
the dotted line shows the case $x_M=x$ and $x_B=1$. In all three cases
the values of $x_M$ and $x_B$ are restricted to the reasonable range of
$0.6 < x <1.1$

\begin{figure}[tbh]
\centerline{\psfig
{%
figure=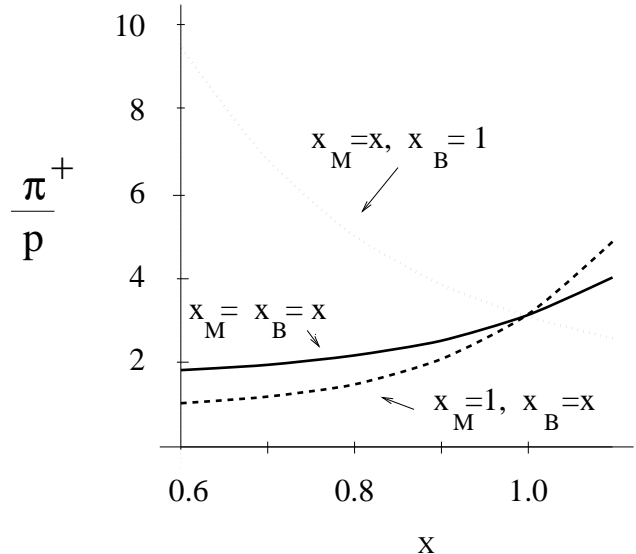,height=11cm,bbllx=44bp,bblly=134bp,bburx=537bp,bbury=724bp,clip=%
}} \label{ps} \vspace{-1.5cm}
\caption{The $\pi ^{+}/p$ ratio plotted as a function of the scaling 
parameter $x$. Solid line: all hadron masses 
(except for Goldstone bosons) are scaled with $x_M=x_B=x$.
Dashed line: only baryon masses are scaled $x_M=1$, $x_B=x$.  
Dotted line: only meson masses (except for Goldstone bosons) are scaled
$x_M=x$, $x_B=1$.
scaled}
\end{figure}

Our results presented in Fig. 1 show that large mass modifications
lead to a substantial change of the $\pi ^{+}/p$ ratio. The
reduction of masses by 20\% changes the $\pi ^{+}/p$ ratio at least by
30\%, which occurs in the case  where $x_M=x_B$. In the case
$x_M=1$ a decrease of baryon masses by 20\% reduces the $\pi ^{+}/p$
ratio by 50\%.  On the other hand, in the case of constant baryon
masses a decrease of meson masses by 20\% enhances the $\pi ^{+}/p$
ratio by 60\%. The displayed behavior can be easily understood in
qualitative terms. The strong increase of the dotted curve in Fig. 1, as the
meson masses are decreased, is mostly caused by a larger population of the
rho and omega mesons at the chemical freeze-out point. Subsequent
decays of these mesons produce more pions, raising the  $\pi ^{+}/p$ ratio.
We note that the rho and omega bring about half of the effect shown
in Fig. 1. The other half comes from heavier resonances.
Similarly, for the dashed curve, a lower mass of baryons results in the 
larger population of protons at the chemical freeze-out point, thus 
lowering the $\pi ^{+}/p$ ratio. A universal scaling of the
meson and baryon  masses
(solid line) partially cancels the above-described effects. Still,
a significant effect remains. 
Other particle ratios are less sensitive to the mass modifications.

To conclude, we note that the thermal fit analysis leaves
little room for modifications of hadron masses larger than,
say, 10-20\%. The quite impressive 
agreement with data reached in \cite{BM3}
would be deteriorated with significant mass modifications.
However, it seems premature to jump to a general conclusion 
that masses of hadrons cannot significantly
change in hot and dense medium. One should bare in mind that
at present we do not understand in sufficient detail the 
mechanisms of hadron production and evolution of the system
created in heavy-ion collisions, and the appealing 
simplicity and numerical success of the thermal model may be misleading. 
One cannot exclude the 
possibility that a more elaborate or altogether 
different treatment will 
leave room for a substantial modification of
hadron masses. 
In order to shed more light on this 
issue it would be useful
to incorporate scaling of hadron masses, as well as other 
hadronic parameters,
in various existing approaches to heavy-ion collisions.

{\it Acknowledgment:}
We thank Krzysztof Golec-Biernat for valuable comments and discussions.


\begin{thebibliography}{9}
\bibitem{BM12}  P. Braun-Munzinger, J. Stachel, J.P. Wessels, and N. Xu,
Phys. Lett. {\bf B344} (1995) 43; Phys. Lett. {\bf B365} (1996) 1

\bibitem{BM3}  P. Braun-Munzinger, I. Heppe, and J. Stachel,  Phys. Lett.
{\bf B} in print, nucl-th/9903010

\bibitem{CR}  J. Cleymans and K. Redlich, Phys. Rev. Lett. {\bf 81} (1998)
5284

\bibitem{Bec} F. Becattini,
Z. Phys. {\bf C69} (1996) 485; J. Phys. {\bf G23} (1997) 1993

\bibitem{H}  U. Heinz, Nucl. Phys. {\bf A638} (1998) 357c;  
CERN-TH/99-209, nucl-th/9907060

\bibitem{Bialas} A. Bialas, Phys. Lett. {\bf B442} (1998) 449;
hep-ph/9909417 

\bibitem{WH} U. A. Wiedemann and U. Heinz, CU-TP-931, nucl-th/9901094

\bibitem{BR}  G. Brown and M. Rho, Phys. Rev. Lett. {\bf 66} (1991) 2720

\bibitem{hatlee}
T. Hatsuda and S.~H. Lee, Phys. Rev. {\bf C46} (1993) R34

\bibitem{hatsuda}
T. Hatsuda, H. Shiomi, and H. Kuwabara, Prog. Theor. Phys. {\bf 95} (1996) 1009

\bibitem{klingl}
F. Klingl, N. Kaiser, and W. Weise, Nucl. Phys. {\bf A624} (1997) 527

\bibitem{cassing}
W. Cassing, W. Ehehalt, and C.~M. Ko, Phys. Lett. {\bf B363} (1995) 35

\bibitem{li}
G.~Q. Li, C.~M. Ko, and G.~E. Brown, Nucl. Phys. {\bf A606} (1996) 568


\bibitem{CERES}  CERES Collab., G. Agakichiev {\it et al.}, Phys. Rev. Lett. 
{\bf 75} (1995) 1272

\bibitem{HELIOS}  HELIOS/3 Collab., M. Masera {\it et al.}, Nucl. Phys. {\bf %
A590} (1995) 3c

\bibitem{Tablice} Eur. Phys. J. {\bf C3} (1998) 25-57

\bibitem{RKH} M. Lutz, S. Klimt and  W. Weise, Nucl.Phys.
{\bf A542} (1992)  521

\end{thebibliography}

\end{document}